# Transformation Invariant Cancerous Tissue Classification Using Spatially Transformed DenseNet

Omar Mahdi, Ali Bou Nassif
*Computing and Informatics*
*University of Sharjah*
Sharjah, UAE
{u16105933, anassif}@sharjah.ac.ae

*Abstract*— In this work, we introduce a spatially transformed DenseNet architecture for transformation invariant classification of cancer tissue. Our architecture increases the accuracy of the base DenseNet architecture while adding the ability to operate in a transformation invariant way while simultaneously being simpler than other models that try to provide some form of invariance.

*Keywords*— Transformers, Spatial Transformers, Deep Neural Networks

## I. INTRODUCTION

One of the most popular and beneficial applications of artificial neural networks (ANNs) is in the field of medical sciences. Image classification and segmentation are common in medical applications, and in recent years, Deep Neural Networks (DNNs) have been increasingly used to carry out vision tasks on medical tasks.

Convolutional Neural Networks (CNNs) have seen great success in image-related tasks and have been widely applied for all kinds of computer vision applications, from the classification of handwritten numbers [1] and objects [2], to COVID-19 classification from X-ray images [3] and attention-deficit hyperactivity disorder (ADHD) diagnosis [4].

One medical application that has seen increased focus recently is applying ANNs to histology images, like tumor classification and nucleus segmentation, both due to advancements in ANN architectures and the availability of more high-quality datasets, like the PatchCamelyon (PCam) [5] and Kumar [6] datasets.

The ANN advancement of interest for this work is the Vision Transformer (ViT). ViTs are based on Transformers, which started as a DNN architecture for Natural Language Processing (NLP) [7], but were later modified to work on image tasks, like object detection [8].

In this paper, we will explore using a spatial transformer network (STN) [9] to improve the performance of the binary classification problem of deciding whether a histological image of a tissue contains a tumor cell or not, by applying DenseNet201 [10] along with a STN frontend for attention on the PCam dataset and using the PyTorch [11] library in the Python programming language.

The rest of the paper is organized as follows: Section II introduces the structure of the spatial transformer network and the convolutional neural network. Section III discusses some work done in the literature about vision transformers and DNN networks for cancer image classification. Furthermore, section IV presents details about the dataset that has been used. Additional experiments and implementation details are given in section V, and finally, the results of the experiments are presented in section VI. The conclusion and future work are discussed in section VII.

## II. TECHNICAL BACKGROUND

### A. Spatial Tranformer Network

The goal of the STN is to get the best input by learning the input's transformation parameters automatically. These parameters are then used to transform inputs into optimal inputs that are then used for feature extraction and classification tasks that follow.

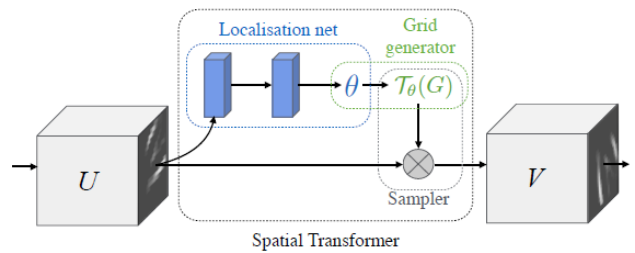

Fig. 1. The architecture of a spatial transformer module. The combination of the localization network and sampling mechanism defines a spatial transformer [9].

The three components of the spatial transformer mechanism are shown in Fig. 1. Initially, the input feature map is routed via a localization network. The spatial transformation parameters are then applied to the feature map through a sequence of hidden layers, resulting in a transformation that is specific to the given input. The projected transformation parameters are then used to construct a sampling grid, which is a collection of locations from which the input map should be sampled to produce the transformed output, which the grid generator does.

Finally, the sampling grid and the feature map are fed into the sampler, which generates an output map sampled at grid points from the input. The STN is divided into three components, which are given below [12].

The first component is the localization network ($f_{loc}$), which employs several hidden layers to generate the transformation matrix $A_\Theta$ from the original input $I$. The parameter form determines the kind of transformation on $I$. In this project, we use 2-D affine transformations, which allow us to do translation, rotation, and scaling transformations with just six parameters.

$$Floc = A_\theta = \begin{bmatrix} a1 & b1 & a3 \\ b2 & a2 & b3 \end{bmatrix} \quad (1)$$

Rotation transformation is provided by a1, a2, b1, and b2, translation transformation is provided by a3 and b3, and scaling transformation is provided by a1 and a2. The second component is the grid generator. It produces sets of points from a sample grid using a regular space grid and the affine transformation matrices $A_\Theta$. These are the locations where $I$ should be sampled in order to provide the transformation result.

$$\begin{bmatrix} u \\ b \end{bmatrix} = A_\theta \begin{bmatrix} x \\ y \\ 1 \end{bmatrix} \quad (2)$$

Where [x, y] represent the $I$ coordinate and [u, v] represent the sample grid coordinates.

The final component is the sampler. $I$ and the sample grid are used as inputs to generate the ideal input $I'$. Because the sampling grid created by the grid generator is not perfectly aligned with the discrete grid values of $I$, the bilinear sampler is used to rectify the pixels in the sampling grid by interpolating from their neighbor pixels.

$$I' = V(p, I) \quad (3)$$

The sampling grid is represented by p, while the bilinear sampler is represented by V.

The ultimate goal of using STN is to achieve high classification accuracy. As a consequence, the used STN, which is a component of the classification system, is part of the end-to-end model. The affine transformation parameters are used to rotate, translate, and scale the original input $I$, yielding the best input $I$ for the succeeding CNN classification network.

*B. Convolutional Neural Network*

The CNN uses the STN network's output images as input images once the STN network has finished the transformation process. A CNN model's three essential components are the convolution layer, the pooling layer, and the nonlinear transformation. The convolution layer's job is to convolve the input (images or feature maps) to generate separate feature maps using various convolution kernels. This is how it is defined:

$$X_l = g(X_{l-1} * W_l + B_l) \quad (4)$$

Where $X_{L-1}$ and $X_L$ represent the input and output of the last convolutional layer, $W_L$ represents the convolution kernels, and $B_L$ represents the biases, the activation function and convolution operation are specified rectified linear units (ReLU) and *, respectively.

When compared to other deep neural network activation functions, the ReLU may be able to avoid difficulties with vanishing gradients and speed up the training process.

By reducing the amount of feature maps, the pooling layer function gives invariance. In most cases, it follows the convolutional layer.

CNN models outperform other DNN models in image classification because they learn the discriminating qualities in a more complicated manner.

The Dense Convolutional Network (DenseNet) [10] has dense connectivity when compared to prior models such as VGG [13] and ResNet [14]. DenseNet solves the vanishing-gradient issue, improves feature map propagation, and reduces the amount of parameters that must be configured.

As illustrated in Fig. 2 below [10], a different connectivity pattern with other CNNs in the DenseNet design incorporates direct connections from any layer to all subsequent levels, which may further improve information flow across layers.

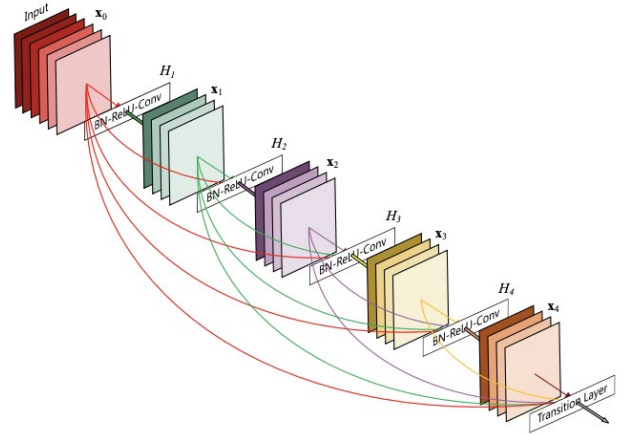

Fig. 2. DenseNet model with a 5-layer dense blocks with a growth rate of k=4 [10].

As a consequence, the l-th layer contains all of the feature maps from the preceding levels. The formula is as follows:

$$x^l = H^l([x^0, x^1, \ldots, x^{l-1}]) \quad (5)$$

Where l represents the layer index and $x^l$ represents the output of the l-th layer. The concatenation of the feature-maps created in layers 0,1,2,...,l-1 is $[x^0, x^1, \ldots, x^{l-1}]$. And $H^l$ may be a composite function of operations like ReLU, Pooling, or Convolution (Conv).

The DenseNet201 model denotes the use of a DenseNet with 201 layers for training [5], as well as changes to input images such as random scaling, rotating, and translating.

## III. RELATED WORK

### A. Vision Transformers Applications

The use of the Transformer method [7] for natural language processing applications has been extremely successful [15]–[17]. Researchers are now investigating the use of the Transformer algorithm for computer vision applications [18]–[20]. However, even though the "vision transformers" technique did very well on the ImageNet dataset [21], it still needs a lot of data to get that level of performance.

Vision Transformer has recently been used for a variety of tasks such as dense prediction, object identification, and image segmentation by deploying pyramid structures and combining them with efficient attention mechanisms [22]–[24] to operate effectively on larger image dimensions [25]–[27]. It has also been used for medical image segmentation [28]–[30] in conjunction with a UNet [31] structure. It is also used in video sequences to distinguish between spatial and temporal attention [32], [33]. Others make changes to the architecture [34], add data and knowledge distillation [35], or use convolutional inductive biases [36] to cut down on the amount of data needed for training.

### B. DenseNet model for Cancer image classification

Patients are generally first subjected to clinical screening, followed by histological study of cancer spots to arrive at a preliminary diagnosis in medical science. Automatic cancer classification utilizing histopathological images is a tough issue for reliably diagnosing cancer, particularly in tiny image patches extracted from larger digital pathology scans. Computer-assisted diagnostics may make this procedure easier while also being more reliable and cost-effective. Image processing has been proposed as a potential technique for disease classification and diagnosis that saves money [37].

A vast amount of research on cancer diagnosis utilizing different image processing and machine learning algorithms has been published [38], [39]. These traditional methods can't be used because they require a lot of time and effort to manually extract traits.

The performance of image classification has lately improved quickly because of the introduction of large-scale hand-labeled data sets (such as ImageNet [40]). Extending deep convolutional networks for medical image classification has received a lot of attention. The VGG16 deep network was used by Wang et al. [41] to detect breast cancer. Esteva et al. [42] did research on skin cancer detection using Inception V3, with the goal of classifying malignancy status. For breast cancer histopathological image classification, Habibzadeh et al. [43] employed the ResNet model. Work has also been done to make it easier to work with histology images that don't change when rotated [44].

Furthermore, research has shown that when compared to traditional machine learning, deep learning approaches are continually used for medical imaging data diagnosis and increase performance [45].

In our work, we will apply the STN as a preprocessing layer to the DenseNet model for the purpose of cancer image classification. Adding the spatial transformer will enhance the dataset by performing the different transformations on the cancer images. More details about the experiments and the results obtained are explained in the next sections.

## IV. DATASET

In this paper, we use the PatchCamelyon (PCam) dataset to test our model and compare it to previous work. The PCam dataset is made up of 327,680 color images, each with a size of 96x96 pixels.

The images are from histopathologic scans of lymph node tissue, and each image has a binary classification of either containing metastatic tissue in the center 32x32 pixel region or not.

The images are pre-divided into training, testing, and verification datasets with 262,144 images for training, 32,768 for testing, and 32,768 for validation. The dataset has an equal split between positive and negative samples, which helps avoid bias when developing and testing the model. This dataset is sent through an STN network that then feeds into a DenseNet201 network pretrained on ImageNet [40].

The pretrained model provided by PyTorch needs 224x224 pixel images and was trained on data with a color mean of [0.485, 0.456, 0.406], and a standard deviation of [0.229, 0.224, 0.225]. Therefore, a normalization preprocessing step is applied to the PCam dataset to match what DenseNet was trained on.

Additionally, to test the model's ability to handle transformed (a combination of translated, rotated, and scaled) inputs, each input image is transformed using a random affine transform, where a rotation of -180≤R≤180 degrees, a translation equal to -0.25≤t1≤0.25 of image width horizontally and -0.25≤t2≤0.25 of image height vertically, and a scale factor of 0.5≤s≤1, are applied.

Examples of the results of such transformations are shown below in Fig. 3.

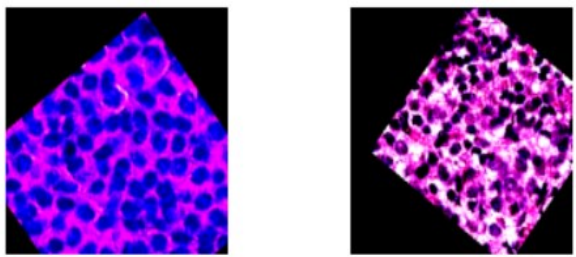

Fig. 3. Examples of transformed input images from PCam dataset.

## V. MODEL

The model chosen for this work is a spatial transform network that starts with a localization network made up of several CNNs and ends with a fully connected layer. The output from the localization network's fully connected layer is fed into a regression network to produce a 3x2 affine transform matrix.

The localization network is made up of 5 convolutional layers, each followed by a 2x2 MaxPool and a ReLU operation. Starting with the 3 color channels as input, the first two convolutional layers in the localization network output 32 channels each, the third and fourth layers output 64 channels, while the fifth layer outputs 128, which is then fed into a regressor network that generates the affine matrix that is used to transform the input.

The generated affine matrix is used to transform the randomly transformed input images such that they help the classification network produce the best results, which can be done, for example, by undoing the random rotation and translation such that the classification network only needs to learn how to perform on one kind of input, instead of having to learn the feature maps of each rotation and translation.

Once the generated affine matrix is applied to the images, these images are fed into the DenseNet201, which is the 201 layers deep classification network. As this is a large network and so needs a lot of data to start performing well, we start with a pretrained DenseNet201 that has been trained on ImageNet.

The localization network, which learns the important parts of the image and provides the attention part of the model, is made up of 5 CNN layers, where each layer is followed by a max pool and then a ReLU function. The final CNN layer's output is flattened and fed into a 2-layer deep fully connected network, which is the regressor network.

The regressor network's first layer is followed by a ReLU, and the final layer outputs 6 values, which are used to build a 3x2 affine transformation matrix. The regression network output layer is initialized with zero weights and biases in such a way to produce an identity affine matrix (one that doesn't affect the input) to ensure that at the start, images are fed as is and transformations are made gradually as the localization network learns what improves the accuracy of the classification network.

## VI. RESULTS

To test our model, we built a comparison between three different models. One model had no STN step, and its inputs were not transformed, which is used as a baseline for a best-case scenario, while the second model had transformed inputs but again without an STN step. Finally, our proposed model has both randomly transformed inputs and an STN step before the classifier.

All models were trained for 20 epochs, where in each epoch the entire training dataset was used, which is ~4000 batches of 64 images each. The results are shown in Figs. 4-6, where the loss is sampled every 100 batches for clarity, and a trendline is shown in red.

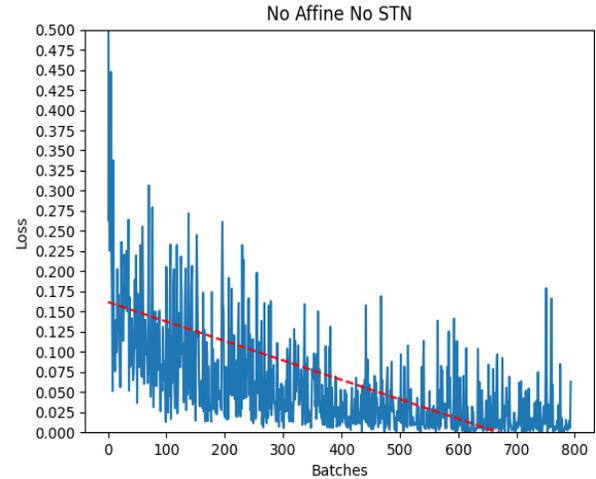

Fig. 4. Loss of 20 epochs for model with standard input and no STN.

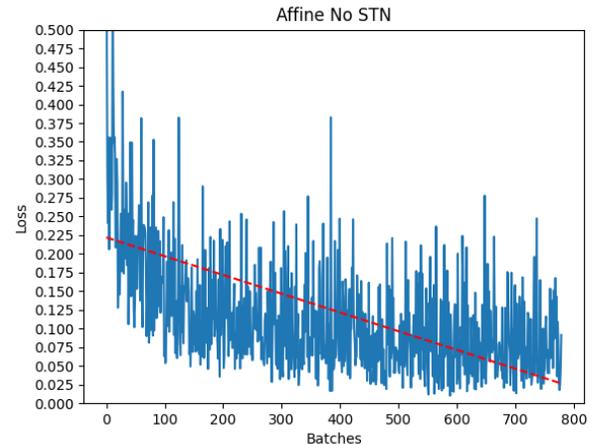

Fig. 5. Loss of 20 epochs for model with transformed input and no STN.

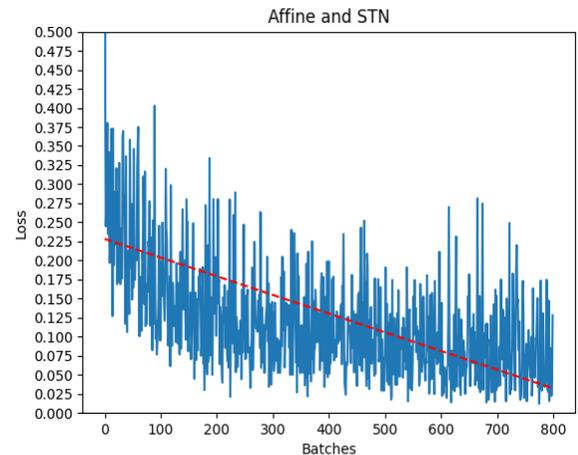

Fig. 6. Loss of 20 epochs for model with transformed input and STN.

It is clear from the first model with normal inputs and no STN that while the loss is spiky, probably due to the size and complexity of the dataset, there is a clear trend down and the model becomes more stable as it is trained.

The effect of transformed inputs is immediately clear on the second model, where the model starts at worse performance, is a lot spikier with big differences between batches, and doesn't reach as good performance in 20 epochs. Additionally, the model in the last few epochs (the last 200 batches in the figure) starts trending up and has worse performance than in the previous epochs.

The third model incorporating the STN starts very similar to the no STN model, which is to be expected as the affine matrix starts at identity, but then produces better performance, especially as training continues. After ~10 epochs, the no STN model continues to see big spikes and then trends up, while with the STN the spikes are less and more consistent, showing less accuracy variation between different inputs, and the model continues trending downwards till the last epoch, showing much better behavior than the pure DenseNet model.

While the STN model does exhibit better characteristics overall, the accuracy itself isn't different in this test, which is an expected effect of introducing the STN network. The downside of adding a transformer is that it introduces a new network that has to be trained from scratch and can take a long time and a lot of data, especially as the complexity of the localization network increases. In this case, the observed better trend of the STN model indicates that with more training, the STN model will continue to improve and eventually produce better results than the no-STN model, as the STN learns to produce better affine matrices. The no STN model, on the other hand, has already shown some signs of plateauing after 20 epochs and is starting to trend towards worse loss values, as it is forced to learn a huge number of feature maps, once per unique transformation, which it will either fail to do or require extremely large amounts of training data.

## VII. Conclusion and Future Work

In this work, we introduced a model incorporating DenseNet201 and a Spatial Transformer Network for cancer classification of color cancerous tissue and showed that models using the STN exhibit better characteristics and continued improvement with training, while the standard DenseNet has large accuracy variance and shows signs of plateau after several epochs.

For future work, we would like to explore the topic further by training for a few hundred epochs and seeing the performance characteristics of the different presented models and by testing different STN networks and seeing which design produces better results.

## VIII. References

[1] F. Siddique, S. Sakib, and M. Abu Bakr Siddique, "Recognition of Handwritten Digit using Convolutional Neural Network in Python with Tensorflow and Comparison of Performance for Various Hidden Layers," in *2019 5th International Conference on Advances in Electrical Engineering (ICAEE)*, 2019, pp. 541–546.

[2] A. Krizhevsky, I. Sutskever, and G. E. Hinton, "ImageNet Classification with Deep Convolutional Neural Networks," *Adv. Neural Inf. Process. Syst.*, vol. 25, 2012.

[3] A. A. Reshi *et al.*, "An Efficient CNN Model for COVID-19 Disease Detection Based on X-Ray Image Classification," *Complexity*, 2021, doi: 10.1155/2021/6621607.

[4] I. Ozsahin and D. Uzun Ozsahin, "Neural network applications in medicine," in *Biomedical Signal Processing and Artificial Intelligence in Healthcare*, Elsevier, 2020, pp. 183–206.

[5] B. S. Veeling *et al.*, "Rotation Equivariant CNNs for Digital Pathology," in *International Conference on Medical image computing and computer-assisted intervention*, 2018, pp. 210–218.

[6] N. Kumar *et al.*, "A Multi-Organ Nucleus Segmentation Challenge," *IEEE Trans. Med. Imaging*, vol. 39, no. 5, pp. 1380–1391, May 2019, doi: 10.1109/TMI.2019.2947628.

[7] A. Vaswani *et al.*, "Attention Is All You Need," *Adv. Neural Inf. Process. Syst.*, vol. 30, Jun. 2017.

[8] N. Carion *et al.*, "End-to-End Object Detection with Transformers," in *European conference on computer vision*, 2020, pp. 213–229.

[9] M. Jaderberg *et al.*, "Spatial Transformer Networks," *Adv. Neural Inf. Process. Syst.*, vol. 28, Jun. 2015.

[10] G. Huang *et al.*, "Densely Connected Convolutional Networks," in *Proceedings of the IEEE conference on computer vision and pattern recognition*, 2017, pp. 4700–4708.

[11] "PyTorch." [Online]. Available: https://pytorch.org/. [Accessed: 21-Nov-2021].

[12] X. He and Y. Chen, "Optimized Input for CNN-Based Hyperspectral Image Classification Using Spatial Transformer Network," *IEEE Geosci. Remote Sens. Lett.*, vol. 16, no. 12, pp. 1884–1888, Dec. 2019, doi: 10.1109/LGRS.2019.2911322.

[13] K. Simonyan and A. Zisserman, "Very Deep Convolutional Networks for Large-Scale Image Recognition," *arXiv Prepr. arXiv1409.1556*, Sep. 2014.

[14] K. He *et al.*, "Deep Residual Learning for Image Recognition," in *2016 IEEE Conference on Computer Vision and Pattern Recognition (CVPR)*, 2016, pp. 770–778, doi: 10.1109/CVPR.2016.90.

[15] Z. Yang *et al.*, "XLNet: Generalized Autoregressive Pretraining for Language Understanding," *Adv. Neural Inf. Process. Syst.*, vol. 32, Jun. 2019.

[16] J. Devlin *et al.*, "BERT: Pre-training of Deep Bidirectional Transformers for Language Understanding," *arXiv Prepr. arXiv1810.04805*, Oct. 2018.

[17] Z. Lan *et al.*, "ALBERT: A Lite BERT for Self-supervised Learning of Language Representations," *arXiv Prepr. arXiv1909.11942*, Sep. 2019.

[18] N. Parmar *et al.*, "Image Transformer," in *International*

*Conference on Machine Learning*, 2018, pp. 4055–4064.

[19] M. Chen *et al.*, "Generative Pretraining from Pixels," in *International Conference on Machine Learning*, 2020, pp. 1691–1703.

[20] K. Han *et al.*, "A Survey on Vision Transformer," *arXiv Prepr. arXiv2012.12556*, Dec. 2020.

[21] O. Russakovsky *et al.*, "ImageNet Large Scale Visual Recognition Challenge," *Int. J. Comput. vision, Springer*, vol. 115, no. 3, pp. 211–252, Sep. 2015.

[22] S. Wang *et al.*, "Linformer: Self-Attention with Linear Complexity," *arXiv Prepr. arXiv2006.04768*, Jun. 2020.

[23] J. Ho *et al.*, "Axial Attention in Multidimensional Transformers," *arXiv Prepr. arXiv1912.12180*, Dec. 2019.

[24] I. Beltagy, M. E. Peters, and A. Cohan, "Longformer: The Long-Document Transformer," *arXiv Prepr. arXiv2004.05150*, Apr. 2020.

[25] Z. Liu *et al.*, "Swin Transformer: Hierarchical Vision Transformer using Shifted Windows," in *Proceedings of the IEEE/CVF International Conference on Computer Vision*, 2021, pp. 10012–10022.

[26] W. Wang *et al.*, "Pyramid Vision Transformer: A Versatile Backbone for Dense Prediction without Convolutions," in *Proceedings of the IEEE/CVF International Conference on Computer Vision*, 2021, pp. 568–578.

[27] P. Zhang *et al.*, "Multi-Scale Vision Longformer: A New Vision Transformer for High-Resolution Image Encoding," in *Proceedings of the IEEE/CVF International Conference on Computer Vision*, 2021, pp. 2998–3008.

[28] J. Chen *et al.*, "TransUNet: Transformers Make Strong Encoders for Medical Image Segmentation," *arXiv Prepr. arXiv2102.04306*, Feb. 2021.

[29] H. Cao *et al.*, "Swin-Unet: Unet-like Pure Transformer for Medical Image Segmentation," *arXiv Prepr. arXiv2105.05537*, May 2021.

[30] J. M. J. Valanarasu *et al.*, "Medical Transformer: Gated Axial-Attention for Medical Image Segmentation," in *International Conference on Medical Image Computing and Computer-Assisted Intervention*, 2021, pp. 36–46.

[31] O. Ronneberger, P. Fischer, and T. Brox, "U-Net: Convolutional Networks for Biomedical Image Segmentation," in *International Conference on Medical image computing and computer-assisted intervention*, 2015, pp. 234–241.

[32] A. Arnab *et al.*, "ViViT: A Video Vision Transformer," in *Proceedings of the IEEE/CVF International Conference on Computer Vision*, 2021, pp. 6836–6846.

[33] G. Bertasius, H. Wang, and L. Torresani, "Is Space-Time Attention All You Need for Video Understanding?," *arXiv Prepr. arXiv2102.05095*, vol. 2, no. 3, p. 4, Feb. 2021.

[34] L. Yuan *et al.*, "Tokens-to-Token ViT: Training Vision Transformers from Scratch on ImageNet," in *Proceedings of the IEEE/CVF International Conference on Computer Vision*, 2021, pp. 558–567.

[35] H. Touvron, M. Cord, M. Douze, F. Massa, A. Sablayrolles, and H. Jégou, "Training data-efficient image transformers & distillation through attention," in *International Conference on Machine Learning*, 2021, pp. 10347–10357.

[36] S. d'Ascoli *et al.*, "ConViT: Improving Vision Transformers with Soft Convolutional Inductive Biases," in *International Conference on Machine Learning*, 2021, pp. 2286–2296.

[37] Z. Zhong *et al.*, "Cancer image classification based on DenseNet model," *J. Phys. Conf. Ser. IOP Publ.*, vol. 1651, no. 1, p. 012143, 2020.

[38] A. Jalalian *et al.*, "Computer-aided detection/diagnosis of breast cancer in mammography and ultrasound: a review," *Clin. Imaging, Elsevier*, vol. 37, no. 3, pp. 420–426, May 2013, doi: 10.1016/j.clinimag.2012.09.024.

[39] P. Darshini Velusamy and P. Karandharaj, "Medical image processing schemes for cancer detection: A survey," in *2014 International Conference on Green Computing Communication and Electrical Engineering (ICGCCEE)*, 2014, pp. 1–6, doi: 10.1109/ICGCCEE.2014.6922267.

[40] J. Deng *et al.*, "ImageNet: A large-scale hierarchical image database," in *2009 IEEE Conference on Computer Vision and Pattern Recognition*, 2009, pp. 248–255, doi: 10.1109/CVPR.2009.5206848.

[41] D. Wang *et al.*, "Deep Learning for Identifying Metastatic Breast Cancer," *arXiv Prepr. arXiv1606.05718*, Jun. 2016.

[42] A. Rezvantalab, H. Safigholi, and S. Karimijeshni, "Dermatologist Level Dermoscopy Skin Cancer Classification Using Different Deep Learning Convolutional Neural Networks Algorithms," *arXiv Prepr. arXiv1810.10348*, Oct. 2018.

[43] M. Jannesari *et al.*, "Breast Cancer Histopathological Image Classification: A Deep Learning Approach," in *2018 IEEE International Conference on Bioinformatics and Biomedicine (BIBM)*, 2018, pp. 2405–2412, doi: 10.1109/BIBM.2018.8621307.

[44] S. Graham, D. Epstein, and N. Rajpoot, "Dense Steerable Filter CNNs for Exploiting Rotational Symmetry in Histology Images," *IEEE Trans. Med. Imaging*, vol. 39, no. 12, pp. 4124–4136, Apr. 2020.

[45] Y. Liu *et al.*, "Detecting Cancer Metastases on Gigapixel Pathology Images," *arXiv Prepr. arXiv1703.02442*, Mar. 2017.